\documentclass[twoside,twocolumn,9pt]{article}
\providecommand{\pgfsyspdfmark}[3]{}
\usepackage{extsizes}
\usepackage[super,sort&compress,comma]{natbib} 
\usepackage[version=3]{mhchem}
\usepackage[left=1.5cm, right=1.5cm, top=1.785cm, bottom=2.0cm]{geometry}
\usepackage{balance}
\usepackage{widetext}
\usepackage{times,mathptmx}
\usepackage{caption}
\usepackage{sectsty}
\usepackage{amsmath}
\usepackage{amssymb}
\usepackage{mwe}
\usepackage{soul}
\usepackage{graphicx} 
\usepackage{lastpage}
\usepackage[format=plain,justification=justified,singlelinecheck=false,font={stretch=1.125,small,sf},labelfont=bf,labelsep=space]{caption}
\usepackage{float}
\usepackage{fancyhdr}
\usepackage{fnpos}
\usepackage[english]{babel}
\usepackage{array}
\usepackage[inline]{enumitem}
\usepackage{droidsans}
\usepackage{charter}
\usepackage[T1]{fontenc}
\usepackage[usenames,dvipsnames]{xcolor}
\usepackage{setspace}
\usepackage[compact]{titlesec}
\usepackage{notes2bib}

\usepackage{bm}
\usepackage{epstopdf}

\definecolor{cream}{RGB}{222,217,201}

\begin{document}

\pagestyle{fancy}
\thispagestyle{plain}
\fancypagestyle{plain}{

\fancyhead[C]{\includegraphics[width=18.5cm]{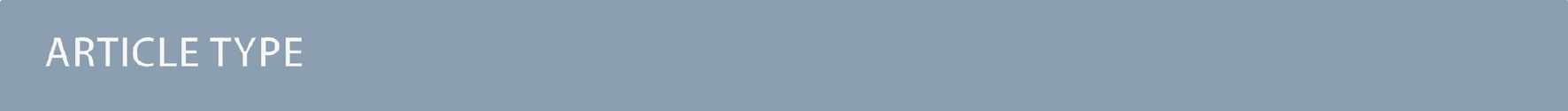}}
\fancyhead[L]{\hspace{0cm}\vspace{1.5cm}\includegraphics[height=30pt]{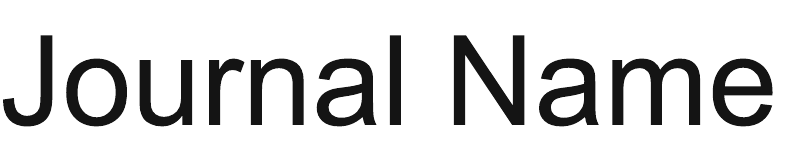}}
\fancyhead[R]{\hspace{0cm}\vspace{1.7cm}\includegraphics[height=55pt]{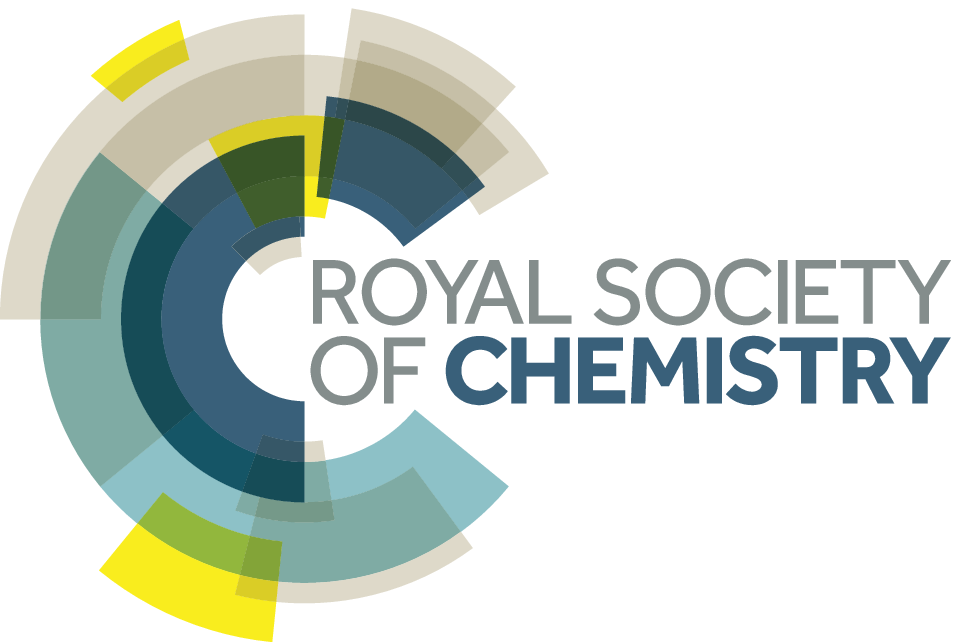}}
\renewcommand{\headrulewidth}{0pt}
}

\makeFNbottom
\makeatletter
\renewcommand\LARGE{\@setfontsize\LARGE{15pt}{17}}
\renewcommand\Large{\@setfontsize\Large{12pt}{14}}
\renewcommand\large{\@setfontsize\large{10pt}{12}}
\renewcommand\footnotesize{\@setfontsize\footnotesize{7pt}{10}}
\makeatother

\renewcommand{\thefootnote}{\fnsymbol{footnote}}
\renewcommand\footnoterule{\vspace*{1pt}%
\color{cream}\hrule width 3.5in height 0.4pt \color{black}\vspace*{5pt}} 
\setcounter{secnumdepth}{5}

\makeatletter 
\renewcommand\@biblabel[1]{#1}            
\renewcommand\@makefntext[1]%
{\noindent\makebox[0pt][r]{\@thefnmark\,}#1}
\makeatother 
\renewcommand{\figurename}{\small{Fig.}~}
\sectionfont{\sffamily\Large}
\subsectionfont{\normalsize}
\subsubsectionfont{\bf}
\setstretch{1.125} 
\setlength{\skip\footins}{0.8cm}
\setlength{\footnotesep}{0.25cm}
\setlength{\jot}{10pt}
\titlespacing*{\section}{0pt}{4pt}{4pt}
\titlespacing*{\subsection}{0pt}{15pt}{1pt}

\fancyfoot{}
\fancyfoot[LO,RE]{\vspace{-7.1pt}\includegraphics[height=9pt]{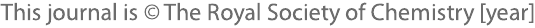}}
\fancyfoot[CO]{\vspace{-7.1pt}\hspace{13.2cm}\includegraphics{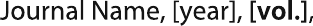}}
\fancyfoot[CE]{\vspace{-7.2pt}\hspace{-14.2cm}\includegraphics{head_foot/RF}}
\fancyfoot[RO]{\footnotesize{\sffamily{1--\pageref{LastPage} ~\textbar  \hspace{2pt}\thepage}}}
\fancyfoot[LE]{\footnotesize{\sffamily{\thepage~\textbar\hspace{3.45cm} 1--\pageref{LastPage}}}}
\fancyhead{}
\renewcommand{\headrulewidth}{0pt} 
\renewcommand{\footrulewidth}{0pt}
\setlength{\arrayrulewidth}{1pt}
\setlength{\columnsep}{6.5mm}
\setlength\bibsep{1pt}

\makeatletter 
\newlength{\figrulesep} 
\setlength{\figrulesep}{0.5\textfloatsep} 

\newcommand{\topfigrule}{\vspace*{-1pt}%
\noindent{\color{cream}\rule[-\figrulesep]{\columnwidth}{1.5pt}} }

\newcommand{\botfigrule}{\vspace*{-2pt}%
\noindent{\color{cream}\rule[\figrulesep]{\columnwidth}{1.5pt}} }

\newcommand{\dblfigrule}{\vspace*{-1pt}%
\noindent{\color{cream}\rule[-\figrulesep]{\textwidth}{1.5pt}} }

\makeatother

\twocolumn[
  \begin{@twocolumnfalse}
\vspace{3cm}
\sffamily
\begin{tabular}{m{4.5cm} p{13.5cm} }

\includegraphics{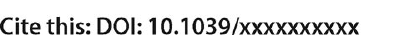} & \noindent\LARGE{\textbf{Manifestation of dipole-induced disorder in self-assembly of ferroelectric and ferromagnetic nanocubes}} \\
\vspace{0.3cm} & \vspace{0.3cm} \\

 & \noindent\large{Dmitry Zablotsky,\textit{$^{a,b}$} Leonid L. Rusevich,\textit{$^{a}$} Guntars Zvejnieks,\textit{$^{a}$} Vladimir Kuzovkov,\textit{$^{a}$} and Eugene Kotomin\textit{$^{a,c}$}} \\

\includegraphics{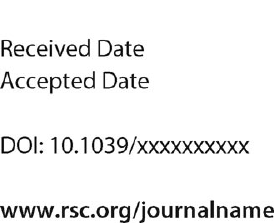} & \noindent\normalsize{The colloidal processing of nearly monodisperse and highly crystalline single-domain ferroelectric or ferromagnetic nanocubes is a promising route to produce superlattice structures for integration into next-generation devices, whereas controlling the local behaviour of nanocrystals is imperative for fabricating highly-ordered assemblies. The current picture of nanoscale polarization in individual nanocrystals suggests a potential presence of a significant dipolar interaction, but its role in the condensation of nanocubes is unknown. We simulate the self-assembly of colloidal dipolar nanocubes under osmotic compression and perform the microstructural characterization of their densified ensembles. Our results indicate that the long-range positional and orientational correlations of perovskite nanocubes are highly sensitive to the presence of dipoles.

} \\

\end{tabular}

 \end{@twocolumnfalse} \vspace{0.6cm}

  ]

\renewcommand*\rmdefault{bch}\normalfont\upshape
\rmfamily
\section*{}
\vspace{-1cm}


\footnotetext{\textit{$^{a}$~Institute of Solid State Physics, Kengaraga str. 8, LV-1063 Riga, Latvia.}}
\footnotetext{\textit{$^{b}$~University of Latvia, Raina bulv. 19, LV-1586 Riga, Latvia. E-mail: dmitrijs.zablockis@lu.lv}}
\footnotetext{\textit{$^{c}$~Max Planck Institute for Solid State Research, Heisenbergstr. 1, 70569 Stuttgart, Germany. E-mail: kotomin@fkf.mpg.de }}


\section{Introduction}
As a fabrication strategy self-assembly\cite{Henzie2011,Quan2012,Singh2014,Caruntu2015,Gao2015,Gong2017} is a promissing pathway to novel materials
 with structural hierarchy, multi-functionality and programmed response to mechanical stress or external field\cite{Jha2012,Zablotsky2017,Feng2017,Zablotsky2019}. It is of great interest to the nanoscale community\cite{Agarwal2011,Damasceno2012,Meijer2017,Wang2018} and fundamental importance for continued technological advancement.
Ferroelectric perovskite-type metal oxides\cite{Varghese2013}, incl. BaTiO\textsubscript{3}, and their solid solutions with SrTiO\textsubscript{3} are high-performance dielectrics widely used in a broad variety of electronic, electro-optic, photovoltaic, electro-chemical\cite{Kuklja2013,Gryaznov2014,Gryaznov2016,Schmid2018}, microelectromechanical (MEMS) components\cite{Acosta2017, Gao2017}. Likewise, efficient ambient or biomechanical energy scavenging\cite{Acosta2017, Gao2015} exploiting piezo-/pyroelectric properties\cite{Krmanc2017,Krmanc2018} of a family of ABO\textsubscript{3} perovskites to power low-energy (incl. wearable or medical) devices is an attractive green energy solution. 
In turn, further downscaling of ferroelectric devices to achieve improved properties (higher dielectric strength, lower loss), which are limited by the grain size, is in strong demand but becoming increasingly difficult in the low nano-range (sub 100 nm) via a conventional top-down paradigm\cite{Acosta2017}. In a different concept a key challenge to further advance the immense practical potential\cite{Kuklja2013} - unique spontaneous polarization and exceptional piezoelectric properties\cite{Rusevich2017, Acosta2017} - of these materials is fabricating highly ordered 3D functional nanoarchitectures (e.g. thin films\cite{Su2016}) over extended areas - ferroelectric supercrystals - from individual ferroelectric nanocrystals\cite{Feng2017}.
Recent advances in chemical synthesis\cite{Adireddy2010,Dang2011,Dang2012,Varghese2013,Caruntu2015} have begun to deliver bulk quantites of nearly monodisperse highly crystalline single-domain nanocubes of some ABO\textsubscript{3} perovskites (incl. BaTiO\textsubscript{3}\cite{Dang2012,Caruntu2015}, SrTiO\textsubscript{3}\cite{Dang2011} and a range of solid solutions), which are ideal space-filling building blocks with a 100\% theoretical packing efficiency, and enabled solution processing via a simple, easily scalable and highly versatile colloidal route to produce superlattice assemblies\cite{Mimura2012b,Caruntu2015, Feng2017} and more complex hierarchies\cite{Mimura2012b,Caruntu2015,Feng2017} for integration into nanostructured ferroelectric devices\cite{Huang2010,Parizi2014,Gao2015}.
Meanwhile, understanding and excercising precise control over the local behaviour of anisometric nanocrystals is absolutely imperative for fabricating highly-ordered assemblies\cite{Mimura2012b}, e.g. for energy harvesting applications\cite{Gao2015,Shin2014}. The current picture of the nanoscale polarization structure\cite{Polking2012, Tyson2014, Sachs2014} and scaling limits of ferroelectric order in individual nanocrystals strongly suggests a potential presence of significant electrodipolar interaction between free-standing ferroelectric monodomains\cite{Yasui2015}. However, the understanding of the role of the dipole-dipole interaction in the condensation of nanocubes beyond the basic scenario of pair-wise attachment\cite{Yasui2015} or the dilute gas-like phase\cite{Zhang2007, Donaldson2015, Donaldson2017, Rossi2018}, in the high-density limit, is an outstanding challenge. 
Here we report the microstructural characterization of densified ensembles produced under osmotic compression from thermalized cubic particles with embedded point dipoles as a minimal model for evaporation-driven self-assembly of ferroelectric and ferromagnetic nanocubes. 
The paper is organized as follows: in Section~\ref{sec:model} we describe our model and the simulation approach: we exploit a discrete element method (DEM) coupled with particle-particle particle-mesh (P\textsuperscript{3}M) approach for computing long-range dipolar interactions; \textit{ab initio} quantum chemical calculations are used for parameter range estimation. In Section~\ref{sec:results} we characterize the produced ensembles employing a variety of statistical descriptors. The experimental implications in relation to ferromagnetic and nanoperovskite nanocube assemblies are discussed in Section~\ref{sec:discussion}. To conclude, Section~\ref{sec:conclusions} provides the summary of this work. 

\section{Methods}\label{sec:model}
\subsection{Self-assembly (DEM) simulations}
In practice a limited amount of self-assembly can be obtained by dropcasting a surfactant-capped nanocube solution onto a compatible substrate (TEM grid or wafer) and drying under ambient conditions \cite{Yu2004, Huang2006, Chang2007, Ren2007, Demortiere2008, Tanaka2011, Choi2012, Eguchi2012, Yang2014, Zhang2016}. However, dropwise deposited assemblies typically lack long-range order and the amount of defects in the produced layers is high owing to a poor control of the growth mode \cite{Agthe2014, Demortiere2008, Henzie2011, Choi2012, Quan2014}. In turn, the immersion methods (e.g. dip coating), whereby the deposition substrate is directly immersed into a nanocrystal solution evaporating under controlled conditions, are much slower (up to several days\cite{Demortiere2008} or weeks\cite{Quan2012}) and afford continuous quasi-equilibrium crystallization into extended 3D superlattices with high degree of coherence over exceptional length scales \cite{Demortiere2008, Zhang2008, Quan2012, Yang2014, Quan2014, Zhang2016} by a controlled evolution of the osmotic pressure at the crystallization front. The assembly of highly crystalline superlattices grown by solution processing is achieved with high-quality nanocrystals having a tight size distribution (standard deviation < 5\%), uniform shape and compatible ligand coverage \cite{ Ren2007, Zhang2008, Quan2012, Zhang2016,  Yang2014}. 
Hence, for self-assembly simulations of monodisperse cubes we use discrete element method (DEM)\cite{Spellings2017} coupled with particle-particle-particle-mesh (P\textsuperscript{3}M) approach for computing long-range dipolar interactions (Supplemental Information). Briefly, the excluded volume interactions between simulated cubes are assembled from a minimal set of lower-dimensional geometric features (i.e. vertices, edges, faces), using point-wise repulsion of all vertex-face and edge-edge pairs at their nearest points. The interaction is modeled by the WCA 12-6 potential ($p=6$), which is a combination of soft potentials
\begin{equation}
  \beta U_{WCA}\left(r_{ij}\right)=4 \left[\left(\delta r_{ij}^{-1}\right)^{2p}-\left(\delta r_{ij}^{-1}\right)^{p}+\frac{1}{4}\right],\quad r_{ij}\leq \delta\sqrt[p]{2} \label{eq:WCA}
\end{equation}
and projects a uniform elastic shell $\frac{1}{2}\delta$ around the cubic shapes reflecting the soft structure of the colloids ligated by an organic coat\bibnote{The typical interparticle spacing in colloid superlattices is $1\div3$ nm as shown by TEM\cite{Choi2012, Quan2014, Caruntu2015, Feng2017}, which is about half as small than twice the length of the surfactant (e.g. $2$ nm for oleic acid/oleylamine) indicating interdigitation or tilt of the covalently bound molecules. Hence, we choose $\delta=0.1\sigma$ for $\sigma$=20 nm.}. 
The dipole-dipole interaction between two assigned point dipoles $\mu_i$ and $\mu_j$
\begin{equation}
  \beta U_{dd}\left(\bm{\mu_{i}},\bm{\mu_{j}},\bm{r_{ij}}\right)=-\frac{\lambda}{r_{ij}^{3}} \left[ 3 \left(\bm{\hat{\mu}_{i}}\cdot\bm{\hat{r}_{ij}}\right)\left(\bm{\hat{\mu}_{j}}\cdot\bm{\hat{r}_{ij}}\right) - \bm{\hat{\mu}_{i}}\cdot\bm{\hat{\mu}_{j}} \right] \label{eq:DD}
\end{equation}
where $\lambda$ is a measure of the relative strength of the dipole-dipole interaction $\propto\lambda k_B T$ vs thermal fluctuations $\propto k_B T$:
\begin{equation}
  \lambda=\frac{1}{4\pi\epsilon_r\epsilon_0}\frac{\beta\mu^2}{\sigma^3} \label{eq:lambda}
\end{equation}
The simulated consolidations of 15 625 cubes starting from low density states were done by slowly ramping the pressure within the isothermal-isobaric (NPT) ensemble using Martyna-Tobias-Klein hybrid scheme (barostat-thermostat) with periodic boundary conditions. The compression runs simulate the entire self-assembly process induced by osmotic pressure during solvent evaporation from a dilute suspension to high density states, which are subsequently measured by statistical descriptors. 

\subsection{\textit{Ab initio} dipole moment calculations}
Structurally both titanium-containing perovskites BaTiO\textsubscript{3} and SrTiO\textsubscript{3} exhibit a cubic (paraelectric) phase above a certain temperature, where each Ti ion is octahedrally coordinated to six oxygen ions. This structure belongs to a centrosymmetric space group, Pm$\overline{3}$m (SG:221), and therefore cannot reveal ferroelectricity, which is a specific property of non-centrosymmetric lattices. In BaTiO\textsubscript{3} the ferroelectric order arises below the Curie temperature (T\textsubscript{C} = 120 $^\circ$C) triggered by a spontaneous tetragonal distortion (P4mm, SG:99) of the prototype cubic symmetry - off-center shift of the Ti\textsuperscript{4+} ions along one of the cube primary axes and a corresponding displacement of ionic sublattices producing a spontaneous polarization (approx. 0.25 C/m\textsuperscript{2} bulk) in the same direction. 
\begin{figure}[h]
\centering
    \includegraphics[width=0.9\linewidth]{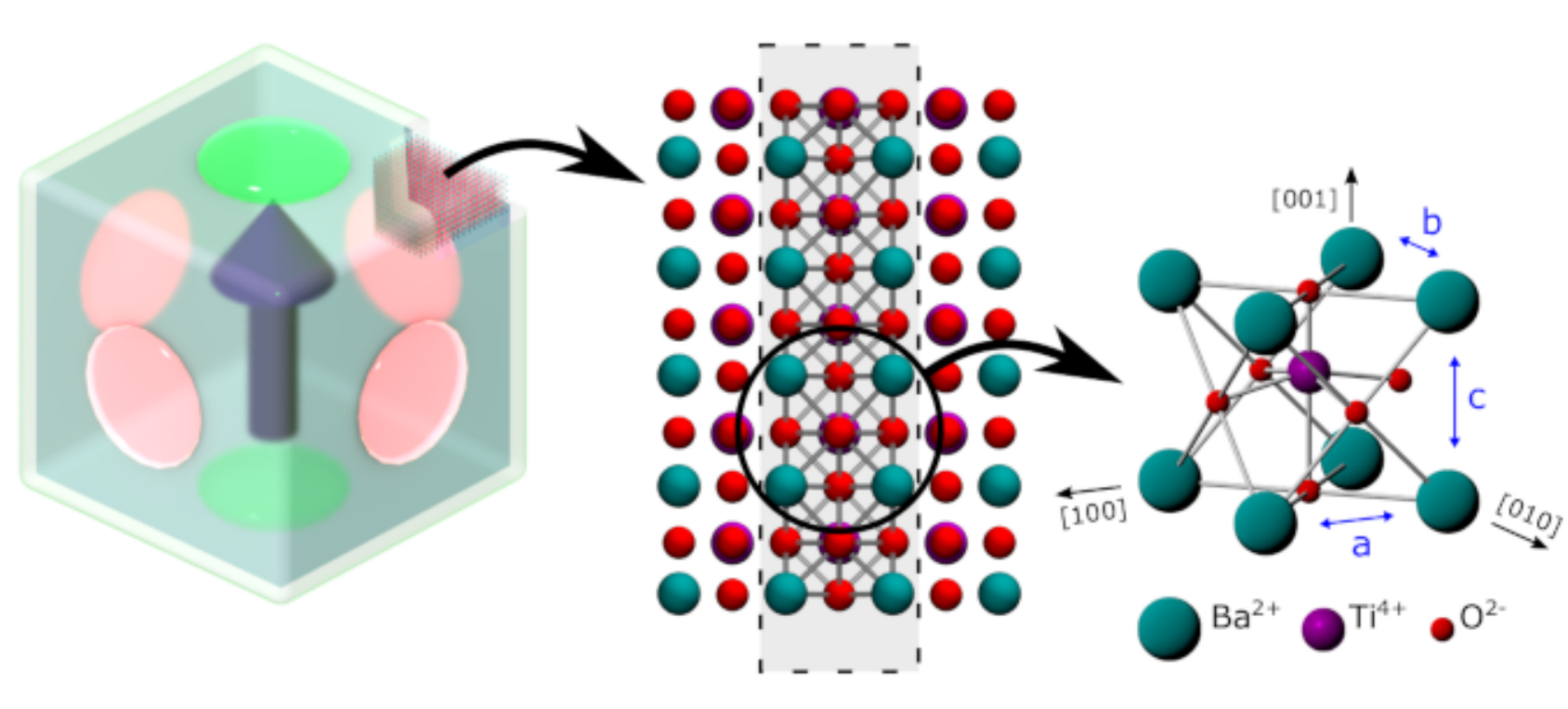}
  \caption{Simulation model: left - model of perovskite nanocubes with dipoles, location of repulsive and attractive patches due to dipole-dipole interaction Eq.~\eqref{eq:DD}; middle - slab structure of perovskite phases used for quantum-chemical calculations (6-12 lattice planes), right - tetragonal distortion of a prototype BaTiO\textsubscript{3} cubic cell (SG:221, $a=b=c$).}
  \label{fgr:tetrag}
\end{figure}
Ferroelectricity evolves via extended cooperative ordering of charge dipoles from local titanium off-center displacements within the Ti-O\textsubscript{6} octahedra. Previous high-energy XRD and atomic PDF studies indicated a progressive decay of ferroelectric coherence across the free standing nanocrystals in the diminished size regime \cite{Petkov2006, Petkov2008, Smith2008, Polking2012, Rabuffetti2012}, whereas the tetragonal distortion itself persists down to sub-10 nm scale \cite{ Petkov2006, Petkov2008, Rabuffetti2012, Caruntu2015}. In turn, high fidelity mapping of ferroelectric structural distortions in discrete sub-10 nm nanocubes of BaTiO\textsubscript{3} via aberration-corrected transmission electron microscopy confirmed a coherent ferroelectric monodomain state with a stable net polar ordering\cite{Polking2012}. Likewise, \textit{in situ} off-axis electron holography showed a clear room-temperature electrostatic fringing field characteristic of a linear dipole emanating from a $\langle001\rangle$ plane \cite{Polking2012} and stable ferroelectric polarization switching.

First-principles (\textit{ab initio}) quantum-chemical estimates of the dipole moments in BaTiO\textsubscript{3} and SrTiO\textsubscript{3} perovskite structure (Supplemental materials) are performed using advanced hybrid functionals of the density-functional-theory (DFT), which already proved reliable for piezoelectric properties\cite{Rusevich2017}; with Hay and Wadt effective small-core pseudopotentials (ECP), previously optimized for BaTiO\textsubscript{3} and SrTiO\textsubscript{3} crystals \cite{Piskunov2004}, for the inner core orbitals of Ba, Sr and Ti atoms and an all-electron basis set for oxygen. The outer-core sub-valence and valence electrons of Ba (5s\textsuperscript{2}, 5p\textsuperscript{6}, 6s\textsuperscript{2}), Sr (4s\textsuperscript{2},4p, \textsuperscript{6},5s\textsuperscript{2}) and Ti (3s\textsuperscript{2}, 3p\textsuperscript{6}, 3d\textsuperscript{2}, 4s\textsuperscript{2}) are calculated self-consistently. A stoichiometric slab structure was cut from a prototype 3D cubic lattice (SG:221, BaTiO\textsubscript{3}: $a=3.993$ {\AA}; SrTiO\textsubscript{3}: $a=3.901$ {\AA}) along $\langle001\rangle$ crystallographic planes. The dipole moment produced by the relaxation of ionic sublattices is used as a lowest estimate for small nanocubes.

\section{Results}\label{sec:results}
\subsection{\textit{Ab initio} parametrization}
A straightforward calculation of the dipole moment for a 20 nm ferroelecric BaTiO$_3$ nanocube based on its bulk polarization density (0.25 C/m\textsuperscript{2}) yields an exceedingly large value $\mu\approx 6\cdot10^{5}$ Debye (D) and $\lambda\approx3.6\cdot10^5$. In turn, the onset of dipole-induced self-assembly in a solution $\lambda_{crit}$ can be estimated from the second $\sim \phi^2 $ virial coefficient of dipolar colloids $b_2=-\frac{e^{2\lambda}}{3\lambda^3}$ (calculated for dipolar spheres\citep{Gennes1970}): from $|b_2|\phi\approx 1 $ for a volume fraction $\phi = 5vol.\%$ (dilute state) $\lambda_{crit}\approx 4$. 
The estimates produced within a phenomenological Landau-Ginzburg-Devonshire theory\cite{Yasui2013}, attempting to account for adsorbate-induced charge screening
, depolarization and potential tetragonal-cubic composite structure,  yielded a dipole moment of approx. 500 D\cite{Yasui2013,Yasui2015} and $\lambda<1$, however, using the bulk parameters for particles of such small size is unreliable. The primary uncontrolled factor here is the degree of adsorbate induced charge screening\cite{Polking2012,Szwarcman2014}.

Likewise, in nanoscale perovskites the near-surface oxygen O\textsuperscript{2-} ions can be slightly displaced with respect to the metal ion of the same plane - surface rumpling\cite{Heifets2000} - known to be quite considerable in many oxide crystals. Here, we have calculated the atomic structure of a periodic slab of cubic BaTiO\textsubscript{3} crystals and optimized the atomic positions in several (varied from 6 to 12, approx. 1-2 nm) $\langle001\rangle$ surface planes using \textit{ab initio} quantum chemical simulations. As a result, we obtain the relaxed slab geometry and characteristic surface dipole moments, which saturate when more than eight lattice planes are allowed to relax, indicative of a large surface polarization and the appearance of a significant electric field near the surfaces of a paraelectric crystal. For the Ti-O\textsubscript{2} termination of the nanocubes (e.g. produced by a hydrothermal route \cite{Crosby2018,Dang2011,Dang2012,Caruntu2015,Su2016}) the surface rumpling of "naked" BaTiO\textsubscript{3} would result in an overall $\langle001\rangle$ dipole of approx. 5300 D and $\lambda=50$, whereas the bulk may still be in the cubic phase. This result is in accord with a previous estimate by some of us using a classical semi-empirical shell model \cite{Heifets2000} (2300 D and $\lambda=10$).
Hence, due to a broad scatter of estimates, in the self-assembly simulations we varied $\lambda\approx 0\div20$ within the indicated range as suggested by our quantum-chemical calculations.

\subsection{Consolidation of simple cubes: reference state}
Colloidal cubes assemble into a  simple cubic (SC) lattice across a broad spectrum of edge lengths from approx. 5 nm to >1 $\mu$m \cite{Yu2004, Ren2007, Demortiere2008, Zhang2008, Chang2007, Henzie2011, Rossi2011, Yang2012, Quan2014}, whereas slight deviations from a cubic shape - truncation or rounding (e.g. due to the adsorbed organic layer modulating excluded volume interactions) - while still maintaining a cubic symmetry, may enable alternative stable assembly pathways with an expanded phase diagram accomodating a rhombohedral (Rh) distortion\cite{Zhang2011, Eguchi2012, Ni2012, Yang2014, Rossi2015, Li2015, Zhang2016, Meijer2017}, body-centred tetragonal (bct)\cite{Quan2014} or face-centered cubic (fcc) \cite{ Quan2012, Choi2012, Demortiere2008} packings as well as their intermediates \cite{Ni2012, Gantapara2013}.

\begin{figure}[h]
\centering
    \includegraphics[width=1\linewidth]{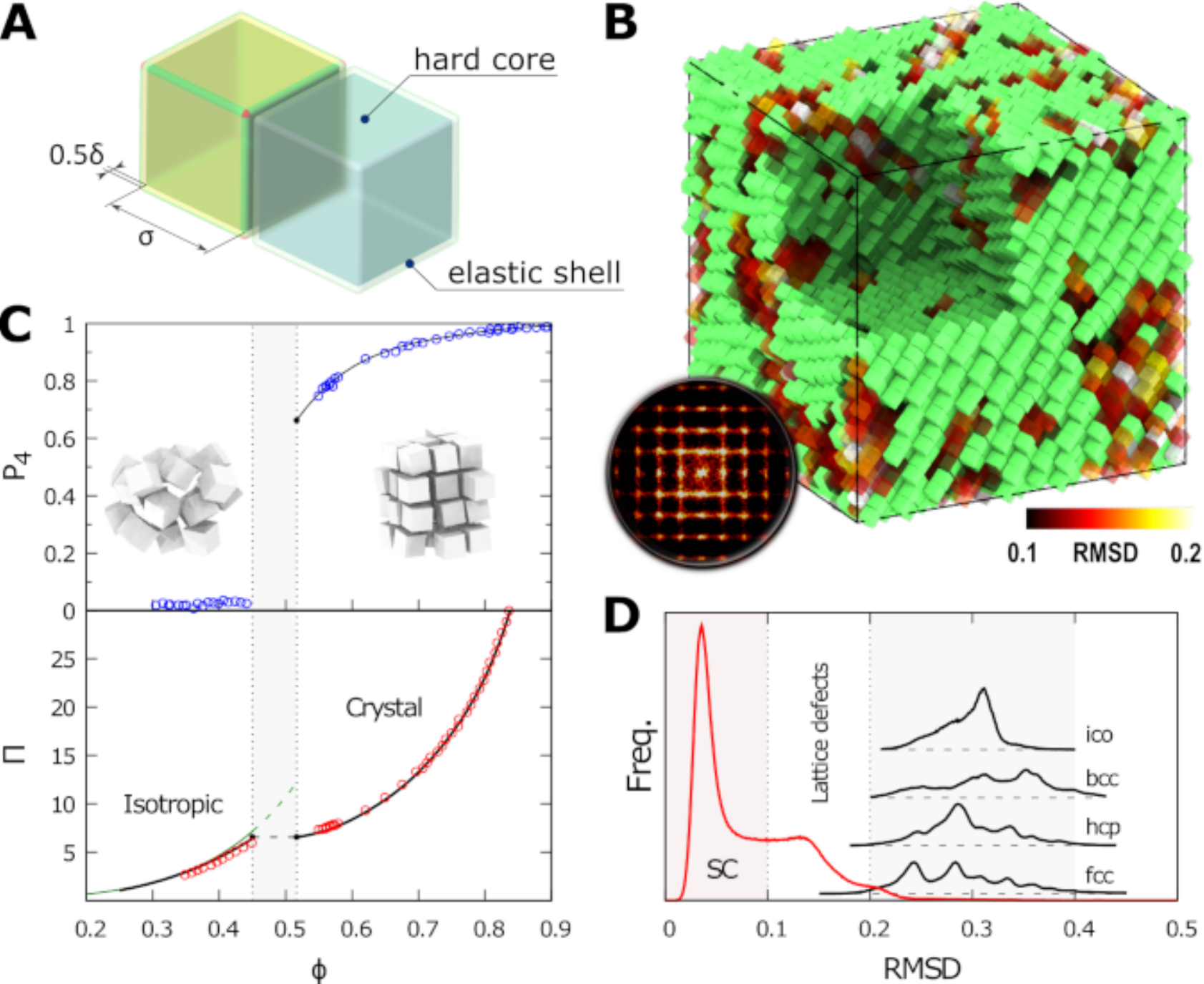}
  \caption{Self-assembly of nanocubes without dipoles as reference state: \textbf{A} - illustration of cube-cube interaction model comprising vertex, edge and face repulsion via a soft elastic shell. \textbf{B} - microstructure of an SC lattice assembled under osmotic compression ($RMSD < 0.1$ in uniform green); inset shows a SAD pattern (illuminated from the high symmetry axis - cubatic director) with characteristic 4-fold symmetry indicating high degree of crystallographic order. \textbf{C} - phase diagram shows the evolution of osmotic pressure $\Pi$ and emergence of crystallographic cubatic order $P_4\rightarrow 1$ with increasing particle volume fraction $\phi$ across a phase transition from dilute isotropic colloid to a crystal state, bold line - Monte-Carlo simulations by Agarwal and Escobedo\cite{Agarwal2011}, dashed line - virial EOS $\Pi=\phi\left(1+\sum_{k=2}^{j}{B_k\phi^{k-1}}\right)$, where $B_2=\left(1+3\alpha\right)$, $\alpha=1.5$, $B_3=18.30341$, $B_4= 41.8485$, $B_5=70.709$, $B_6=88.33$, $B_7=63.5$, $B_8=-37$ according to Irrgang et al.\cite{Irrgang2017}. \textbf{D} - deviation of local structure (RMSD spectrum) from a set of common lattices for the assembly shown in \textbf{B} reliably identifies SC lattice as the dominant motif.}
  \label{fgr:l0ptm}
\end{figure}
To establish a reliable reference state we start by simulating the assembly of simple cubes without dipoles (Fig.~\ref{fgr:l0ptm}A, B). Under osmotic compression the cubes readily form a simple cubic (SC) crystal as the ensemble density is increased.
The transition between a dilute state and a crystalline solid is first order \cite{Smallenburg2012} with the coexistence packing densities $\phi$ between 0.45 and 0.5-0.52\cite{Agarwal2011,Smallenburg2012}, slightly modulated by the amount of delocalized vacancies in the system\cite{Smallenburg2012}. The equation-of-state (EOS) as a function of packing density (Fig.~\ref{fgr:l0ptm}C) compares well to the previous Monte-Carlo simulations \cite{Agarwal2011} of hard cubes. For the isotropic state the correspondence with a perturbative virial EOS\cite{Irrgang2017} expressed as a power series in density is reasonable as well. 
The long-range positional and orientational symmetries emerge robustly at high density: cubatic order parameter\cite{Agarwal2011} $P_4\in \left[0,1\right]$ (Supplemental Information) (Fig.~\ref{fgr:l0ptm}C), which is sensitive to the symmetries of the cubic lattice, indicates the emergence of global rotational correlations. The virtual selected area (electron) diffraction\cite{Coleman2013} (SAD) (irradiated along the cubatic director) (Fig.~\ref{fgr:l0ptm}B, inset) probes the long-range order and shows a characteristic 4-fold symmetric spot pattern corresponding to the expected crystallographic orientations. Likewise, polyhedral template matching\cite{Larsen2016,Stukowski2009} (PTM) is used to assign the crystalline structure by graph-based matching of the convex hull formed by the local neighborhood to a set of predetermined templates (Fig.~\ref{fgr:l0ptm}B). The spectrum of the root-mean-square deviation (RMSD, Fig.~\ref{fgr:l0ptm}D) for a set of standard lattices reliably identifies the SC phase as the dominant motif and its structural distortions. The spectrum is sharp and well separated. The extended shoulder of the SC peak is assigned to the volume conserving finite temperature row-displacements (shearing) induced by collective thermal vibrations accommodated within the monocrystal. After manually inspecting the configurations, we have not observed a single jammed state in the assembled SC lattices, which shows that the kinetic effects are minimal and the compression protocol is adequate. 

\subsection{Self-assembly of dipolar nanocubes}
After assigning $\langle001\rangle$-dipoles to the initial dilute ensemble the compression protocol is repeated with varying $\lambda$-parameter to access a range of magnitudes of the dipolar interactions, after which a series of metrics are applied to the consolidated state to assess its structural properties.
\begin{figure}[ht]
\centering
    \includegraphics[width=1\linewidth]{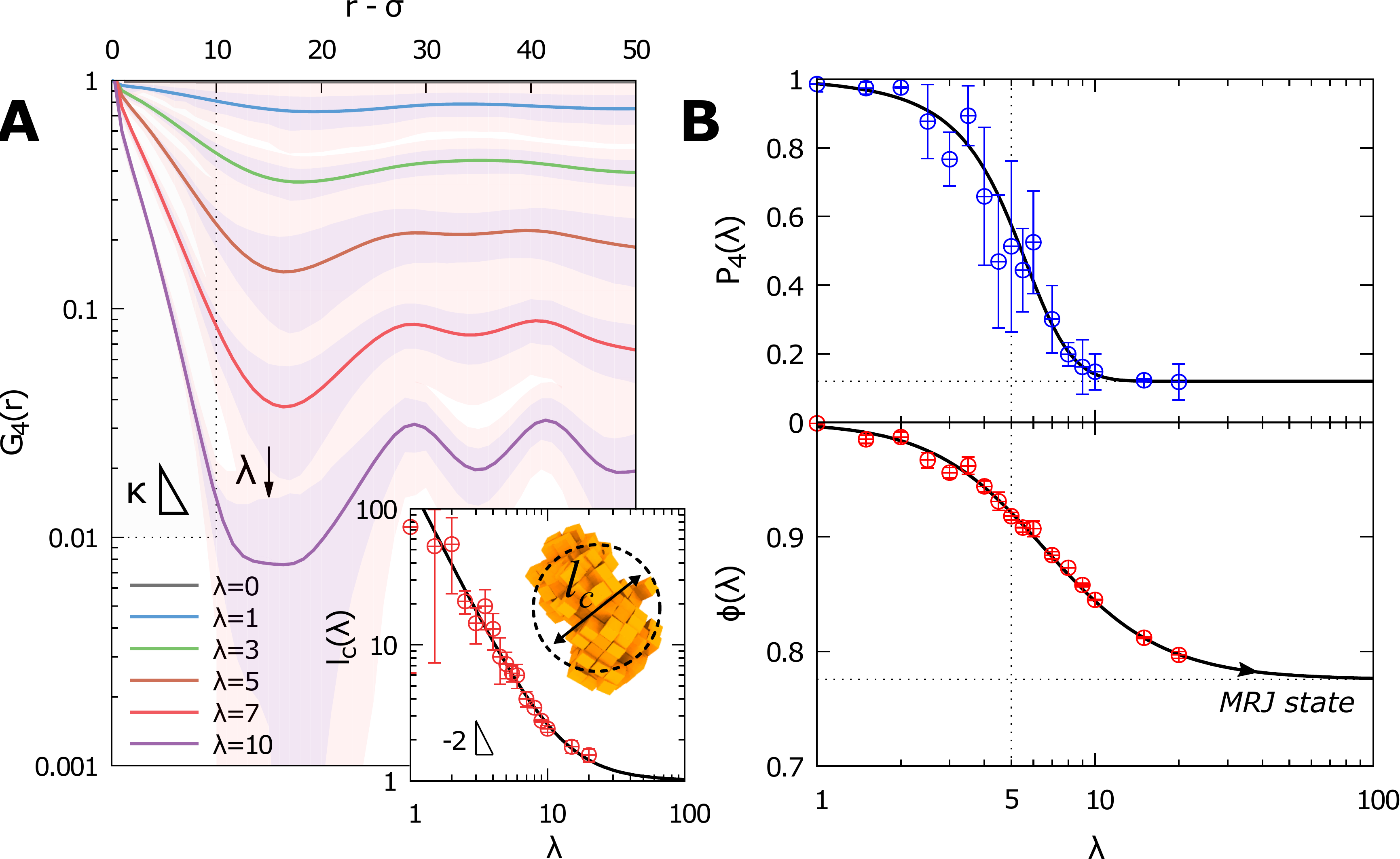}
  \caption{Mesoscale structural correlations in self-assembled solids: \textbf{A} - cubatic orientational correlation function $G_4\left(r\right)$ indicates a progressing polycrystallinity and the decay of cubatic correlation length $l_c=\kappa^{-1}$ (inset) assigned to the shrinking of crystallite size with increasing strength of nanocube dipoles $\lambda$. \textbf{B} - loss of global cubatic order $P_4$ and decrease of packing fraction $\phi$ (enhanced porosity) at higher $\lambda$. The consolidated packings become MRJ states $\phi\rightarrow 0.77$ at high interaction strength $\lambda\gg1$.}
  \label{fgr:g4map}
\end{figure}
We measure the positional extent of the mesoscale cubatic correlations using the orientational correlation function $G_4\left(r\right)\in\left[0,1\right]$ (Supplemental Information), which quantifies the degree of mutual alignment of cubes as a function of their separation distance $r$. For assemblies of simple cubes ($\lambda=0$) forming an SC lattice $G_4=1$ irrespective of distance indicating the presence of strong long-ranged correlations extending beyond the sampling ability of the simulation box. For all $\lambda>0$ (Fig.~\ref{fgr:g4map}A) $G_4\left(r\right)\propto e^{-\kappa\left(r-\sigma\right)}$ shows an exponential loss of orientational correlation in the immediate neighborhood ($r\lesssim 10\sigma$). The characteristic length of the correlated region $l_c=\kappa^{-1}$ is a power law with respect to the interaction parameter $l_c\propto \lambda^{-a}$ (where $a\approx 2$) and quickly drops to just about a few particle sizes. For $\lambda\gg 1$ it appears that $l_c\rightarrow 1$ consistent with the fact that the correlation length cannot be smaller than approx. a single particle size $\sigma$. 

Likewise, the cubatic order parameter $P_4$ (Fig.~\ref{fgr:g4map}B) measured for the whole ensemble displays a progressive diminishment of the global 4-fold alignment with increasing strength of dipolar interactions. Since the asymptotic behavior of $G_4\left(r\right)\rightarrow P_4^2$ as $r\rightarrow\infty$ and the remaining global cubatic order as $\lambda \rightarrow \infty$ is $P_4\approx 0.1$, the tail of $G_4$ cannot decrease beyond $\propto10^{-2}$, which is consistently observed in simulations. At the same time the decrease of the packing fraction $\phi$ in the consolidated samples constitutes a deviation from the shape-optimal packing behavior and the abundance of mechanically jammed states at high interaction strength $\lambda\gg 1$. The maximally random jammed (MRJ) packings of monodispersed cubes, defined as the jammed state with a minimal value of an order metric \cite{Torquato2010}, were shown to have relatively robust densities of approx. 0.77 \cite{Torquato2010, Liu2017}. Hence, the consolidated packings become MRJ states at high interaction strength. 

\begin{figure}[ht]
\centering
    \includegraphics[width=\linewidth]{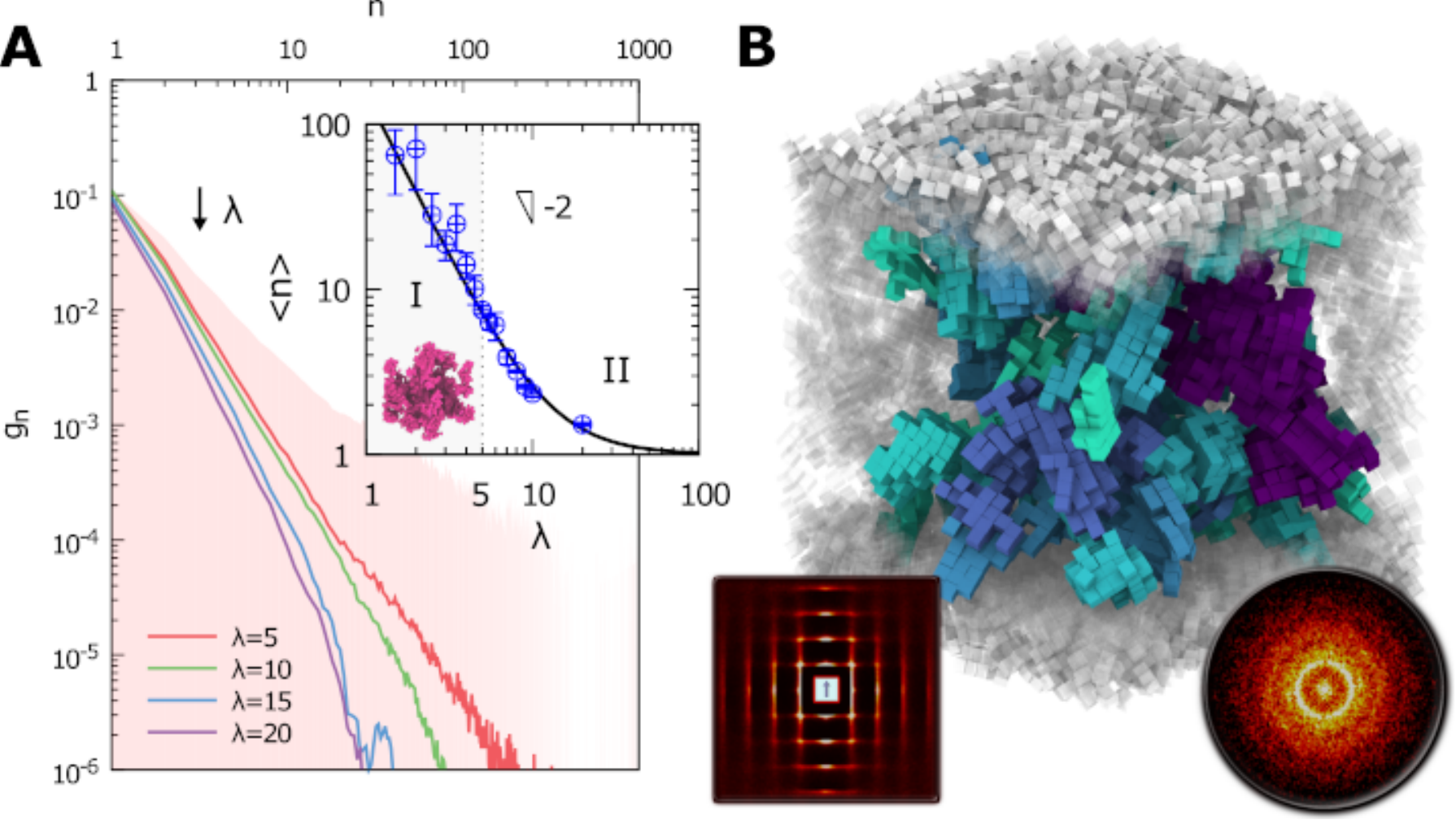}
  \caption{Crystallite size analysis in the assembled solids: \textbf{A} - cluster size distribution $g_n$ (number of $n$-particle clusters with cubic lattice per unit volume) at varying interaction strength $\lambda$, inset shows a decaying average crystallite size $\langle n\rangle$ as a function of $\lambda$ ($I$ - a single percolating cluster exists in the system, $II$ - crystallites are embedded within a glassy solid and well separated). \textbf{B} - microstructure of an ensemble densified at $\lambda=7$ showing a collection of 50 largest crystallites, left inset - pair probability distribution $g_2\left(\mathbf{r}\right)$ ($g_2\left(\mathbf{r}\right)dV$ is the probability of finding a nanocube in the vicinity of point $\bm{r}$ relative to a reference nanocube) shows a well structured local neighbourhood within individual crystallites, right inset - virtual SAD ring pattern (irradiated along the cubatic director) indicates the local coordination of nanocubes and the absence of long-range order. }
  \label{fgr:gn_lambda}
\end{figure}

The decay of both a global order $P_4$ and the correlation range $l_c$ is compatible with a picture of progressing policristallinity of an initially monocrystalline ensemble. To further characterize this structural transformation we perform a cluster size analysis using a proximity-based clustering criterion\cite{Zablotsky2017} - cubes with separation $r<\sigma_c$, where $\sigma_c=1.15$ encompasses just the first coordination sphere of tightly bound (face-to-face) cubes, are considered to be a part of the same cluster - in conjunction with PTM to identify local nuclei (i.e. embedded crystallites) with roughly SC symmetry (Fig.~\ref{fgr:gn_lambda}B). The calculated cluster size distribution $g_n$ (Fig.~\ref{fgr:gn_lambda}A) represents the number of $n$-particle clusters per unit volume (normalized by particle volume $\sigma^{-3}$) assembled under the specific conditions of the dipolar interaction strength $\lambda$ and shows a decaying power-law dependence with respect to the cluster size $g_n\propto n^{-b}$. The total number of crystallites decreases ($g_n$ decreases in magnitude) and the size of the crystallites diminishes ($g_n$ shifts towards lower $n$) with stronger dipolar interactions.
Fig.~\ref{fgr:gn_lambda}B illustrates a few ($50$) of the largest clusters. Visual inspection indicates that for $\lambda\lesssim 5$ a percolating cluster exists in the system. The ensemble is best described as a polycrystalline solid, whereas for $\lambda\gtrsim 5$ we observe a collection of well separated single-crystalline nuclei embedded in a glassy-like matrix. The characteristic size of the clusters is estimated 
\begin{equation}
  \langle n\rangle=\frac{\sum_n n g_n}{\sum_n g_n}
\label{eq:n_avg}
\end{equation}
The average cluster size is consistent with the correlation range analysis, showing a similar behavior and magnitude (cf. Fig.~\ref{fgr:g4map}A, inset).

\begin{figure}[ht]
\centering
    \includegraphics[width=1\linewidth]{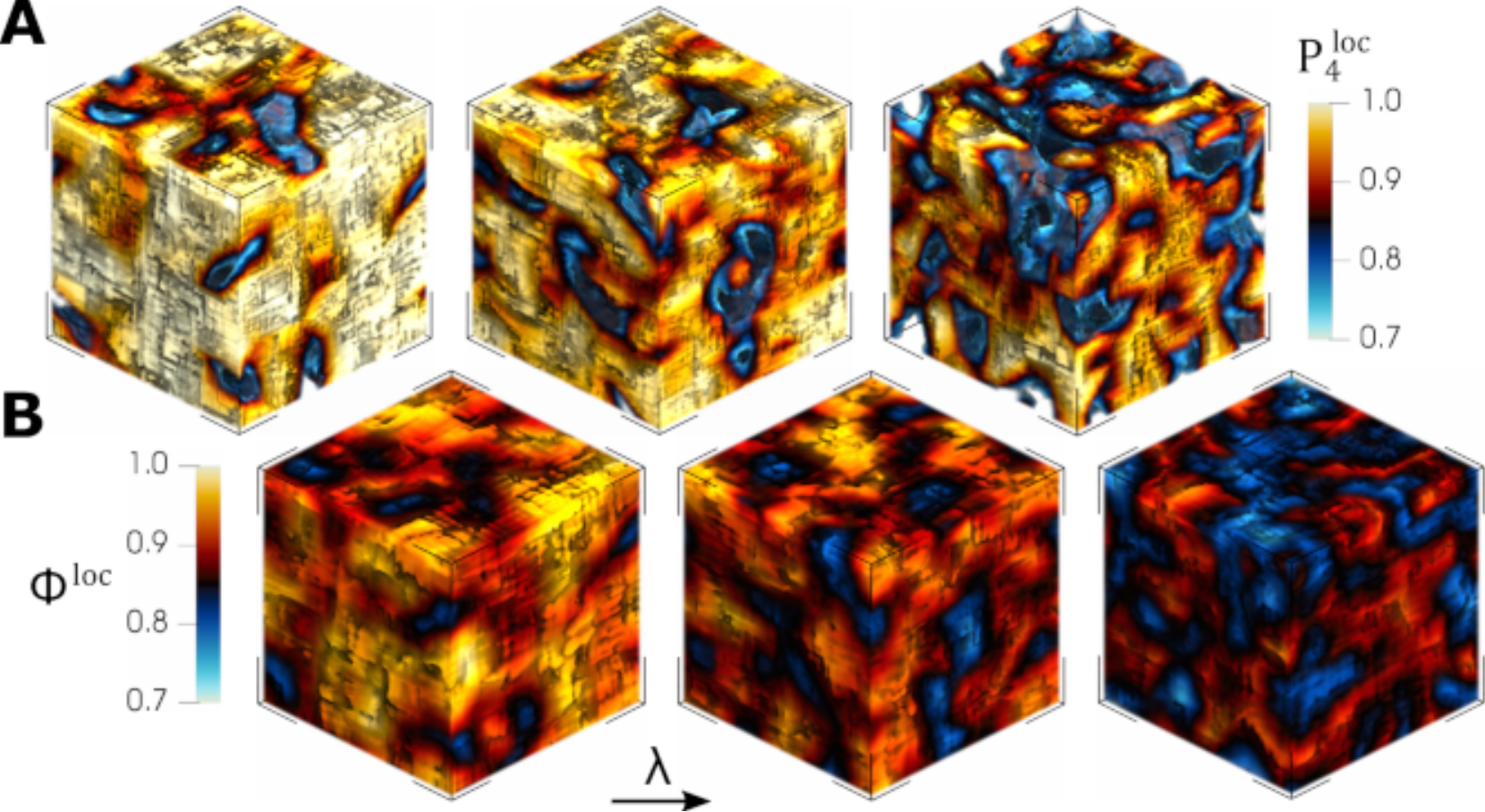}
  \caption{Topography of mesoscale positional and orientational correlations in densified ensembles ($\lambda=3,5,7$, left to right): \textbf{A} - 3D map of local cubatic order parameter $P_4^{loc}$ and \textbf{B} - distribution of packing fraction $\Phi^{loc}$ within the simulated domain. Overlap indicates the location of embedded crystallites. }
  \label{fgr:3dmaps}
\end{figure}

In turn, to examine the locality of orientational and positional correlations we produce detailed topographic maps (Fig.~\ref{fgr:3dmaps}) of the local cubatic order parameter \cite{Liu2017} and packing density fluctuations (Supplemental materials) calculated over the local neighbourhoods bounded by the first 2 coordination shells in the consolidated assemblies. 
These mappings indicate progressing loss of crystallinity and fragmentation of the initial monocrystal with increasing magnitude of dipolar interactions in accord with the analysis based on the ensemble-averaged descriptors. The overlap between the locality of the cubatic order $P_4^{loc}$ and packing fraction $\Phi^{loc}$ marks inclusions with both a high positional and orientational order of the simple cubic lattice, separated by an interface with reduced order indicative of polycrystallinity. The dipole-induced formation of a glassy structure and porous space within the layer is an undesirable complication for the rational design of ferroelectric devices\cite{Feng2017}, e.g. via strain engineering. 

\subsection{Mechanism of order frustration}
The ordering of densified ensembles is triggered by the thermodynamic preference of crowded colloids to minimize free energy by optimizing their local packing \cite{vanAnders2013, vanAnders2014, Damasceno2012, Baule2018} under osmotic compression; whereby the directionality of entropic forces controlling the crystallization of hard colloids statistically emerges around the geometrical features from the collective tendency of the ensemble to increase entropy by osmotic self-depletion.
We compute the effective pairwise potential of mean force and torque (PMFT) to quantify the emergent entropic valence driving the formation of the locally aligned crystalline neighborhoods and identify the enthalpic effect of added dipolar interactions, which are treated within a common context by the PMFT, derived by extracting reference coordinate pair $\Delta\xi_{12}$ (position and orientation) and treating the rest implicitly\cite{vanAnders2013,vanAnders2014}
\begin{equation}
  \beta F_2\left(\Delta\xi_{12}\right)=\beta U\left(\Delta\xi_{12}\right)-\log{J\left(\Delta\xi_{12}\right)}+\beta \tilde{F}_2\left(\Delta\xi_{12}\right)
\label{eq:pmft5}
\end{equation}
Here $U\left(\Delta\xi_{12}\right)$ is the pairwise interaction potential, i.e. steric (Eq.~\eqref{eq:WCA}) and dipolar (Eq.~\eqref{eq:DD}), the Jacobian $J\left(\Delta\xi_{12}\right)$ encodes the entropy of the reference pair accounting for the number of ways the relative position and orientation $\Delta\xi_{12}$ can be assumed, whereas $\tilde{F}_2\left(\Delta\xi_{12}\right)$ is the remaining free energy of the ensemble with the reference pair fixed. The latter terms can be viewed as providing finite density correction to $U\left(\Delta\xi_{12}\right)$. We calculate the 3D topology of PMFT for both simple cubes (as reference) and dipolar cubes ($\lambda=7$) from the pair distribution function\cite{vanAnders2013,vanAnders2014}:
\begin{equation}
  \beta F_2\left(\bm{r}\right)=-\ln{g_2\left(\bm{r}\right)}+const.
\label{eq:pmft}
\end{equation}
The involved statistical integrals are computed from an additional series of NVT trajectories branching from the NPT compression runs after reaching the desired density.
\begin{figure}[ht]
\centering
    \includegraphics[width=0.9\linewidth]{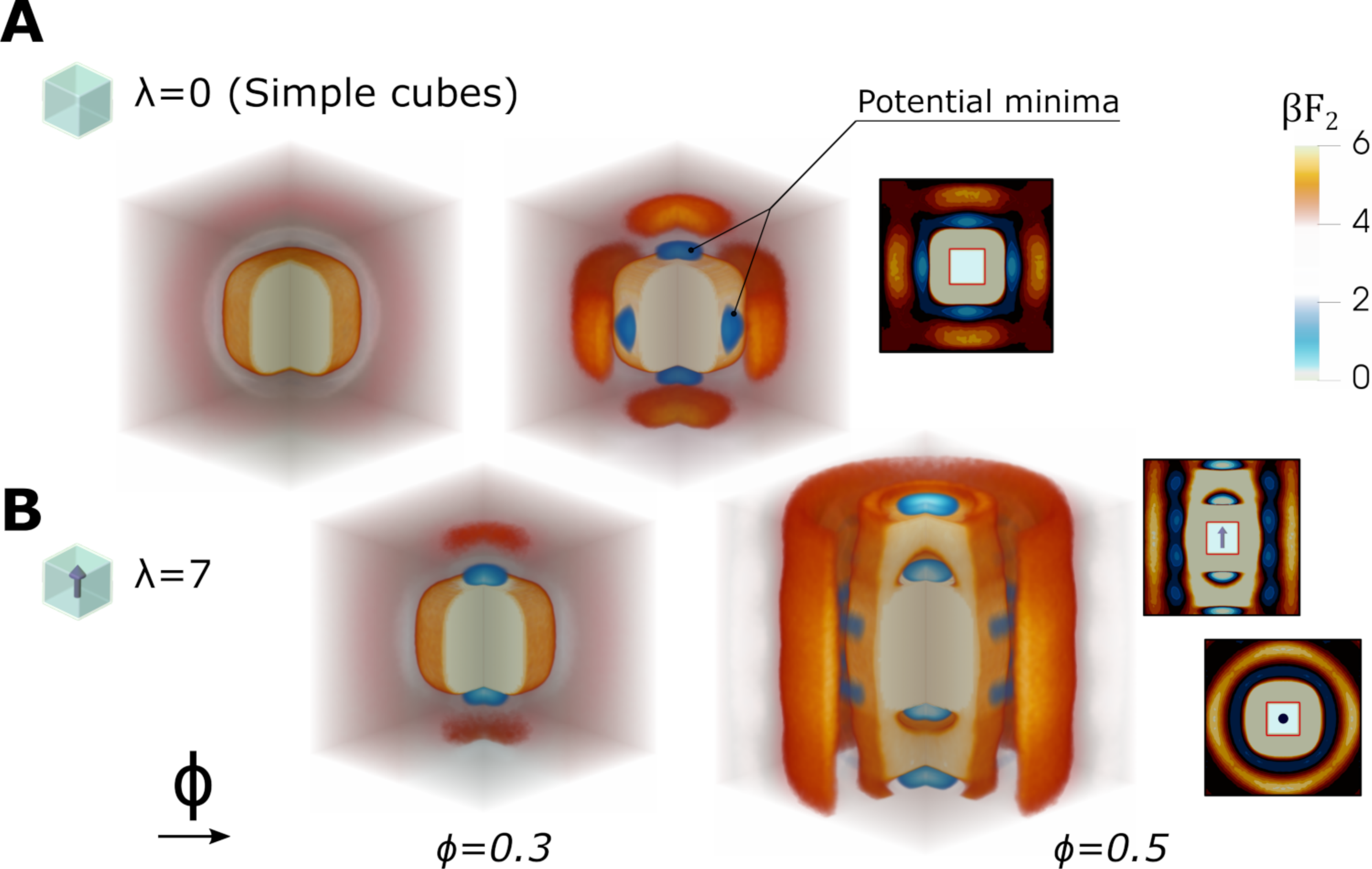}
  \caption{The local distribution of potential of mean force and torque (PMFT) of nanocubes (up to an additive constant) absorbing collective interactions at varying colloid volume fraction $\phi=0.3$ (isotropic state) and $0.5$ (onset of crystallization): for \textbf{A} - simple cubes and \textbf{B} - cubes bearing $\langle001\rangle$ dipoles ($\lambda=7$). }
  \label{fgr:pmft}
\end{figure}
In the absence of dipole-induced enthalpic attraction in $U\left(\Delta\xi_{12}\right)$ the self-assembly is a competition between the osmotic self-depletion $\tilde{F}_2\left(\Delta\xi_{12}\right)$, produced by the ensemble on the reference pair, and the tendency of the pair to assume local dense packing configurations modulated by the particle shape (via excluded volume), in order to minimize the overall free energy $\beta F_2\left(\Delta\xi_{12}\right)$. The PMFT (Fig.~\ref{fgr:pmft} becomes anisotropic as the particle volume fraction is increased, the potential minima $\propto kT$ appear gradually and become sharper and narrower at higher $\phi$. For simple cubes the primary wells of emerging entropic valence are located at octahedral sites coordinating to the cube's facets imposing high positional and orientational correlation at these locations to eventually produce face-to-face alignment. In turn, the logarithmic relationship \eqref{eq:pmft} indicates that a relatively small change in the free energy, such as an enthalpic contribution to the pairwise interaction $ U\left(\Delta\xi_{12}\right) $, may significantly influence both the local coordination environment $g_2(\bm{r})$ of the low density state and the crystallography of the assembled solid. For dipolar cubes the added interaction produces coordination at $\langle 0,0,\pm1\rangle$ sites already in solution associated with the chain fluid state\cite{Zablotsky2017, Zablotsky2019}. Further crowding blocks these sites by potential barriers of the order of $\propto \lambda kT$, which are rather impermeable and restrict the diffusion of the cubes within the local neighborhood leading to geometric frustration during the annealing phase. 

\section{Discussion}\label{sec:discussion}

\subsection{Ferromagnetic nanocubes (Co\textsubscript{x}Fe\textsubscript{3-x}O\textsubscript{4})}

\begin{figure*}[h]
\centering
    \includegraphics[width=1\linewidth]{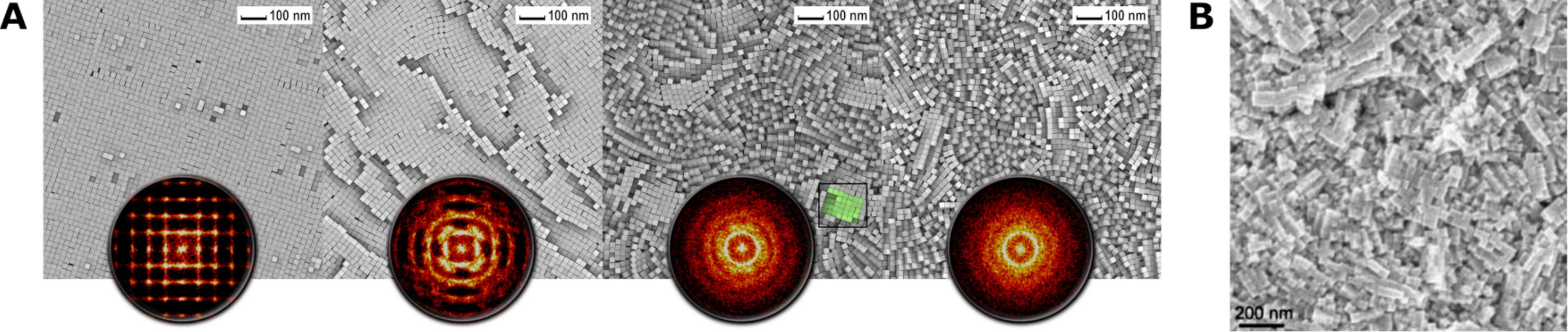}
  \caption{Simulated surface morphology of layers assembled under varying strength of dipolar interaction $\lambda$ displaying signature features of dipole-induced disorder: \textbf{A}, left to right - simple cubes ($\lambda=0$) forming regular lattice, dipolar cubes ($\lambda=2,5,10$) self-assemble into several distinct crystallites embedded into a mostly random structure. Virtual SAD patterns confirm the decay of long-range order. Marked area indicates the location of the crystallite fragment shown in detail in Fig.~\ref{fgr:spin}. \textbf{B} - SEM image of approx. 35 nm ferromagnetic Co\textsubscript{x}Fe\textsubscript{3-x}O\textsubscript{4} ($\lambda\gg 1$) nanocubes self-assembled on a Si substrate [reproduced from Ref.~\cite{Yang2015} with permission from The Royal Society of Chemistry].}
  \label{fgr:simul}
\end{figure*}

\begin{figure*}[!ht]
\centering
    \includegraphics[width=1\linewidth]{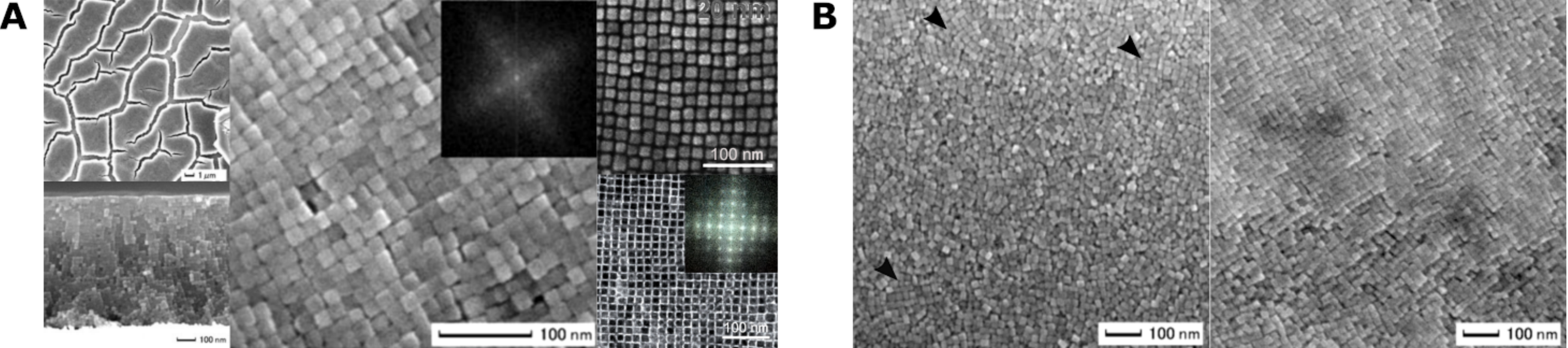}
  \caption{\textbf{A} - Microstructure of ordered BaTiO\textsubscript{3} films on a Pt/MgO substrate (in-plane and crossection, left) and magnified image (middle, inset shows the corresponding FFT pattern), Ref.~\cite{Mimura2013} [reprinted by permission from Springer Nature: Springer Nature,  J. Nanoparticle Res. (15), Mimura, K. \& Kato, K., Fabrication and piezoresponse properties of $\{100\}$ BaTiO3 films containing highly ordered nanocube assemblies on various substrates, Copyright (2013)]; long-range-ordered superlattice structures of BaTiO\textsubscript{3} nanocubes produced by capillary bridge manipulation (right, inset shows the corresponding wide-angle SAED pattern, in which a 4-fold symmetry is clearly observed, indicating the formation of a cubic-type superlattice structure), Ref.~\cite{Feng2017} [from Adv. Mater. (29), Feng, J. et al., Large-Scale, Long-Range-Ordered Patterning of Nanocrystals via Capillary-Bridge Manipulation, Copyright (2017), reprinted by permission of John Wiley \& Sons, Inc]. \textbf{B} - SEM images of 20 nm SrTiO\textsubscript{3} nanocube assembly and BaTiO\textsubscript{3}-SrTiO\textsubscript{3} 1:1 nanocube mixture assembly fabricated by dip-coating, Ref.~\cite{Mimura2012b} [reproduced from Appl. Phys. Lett. (101), Mimura, K. et al., Piezoresponse properties of orderly assemblies of BaTiO\textsubscript{3} and SrTiO\textsubscript{3} nanocube single crystals, Copyright (2012), with the permission of AIP Publishing], with arrows we have marked the location of a few identifiable embedded crystallites with SC lattice. }
  \label{fgr:exper}
\end{figure*}

\begin{figure}[h]
\centering
    \includegraphics[width=0.9\linewidth]{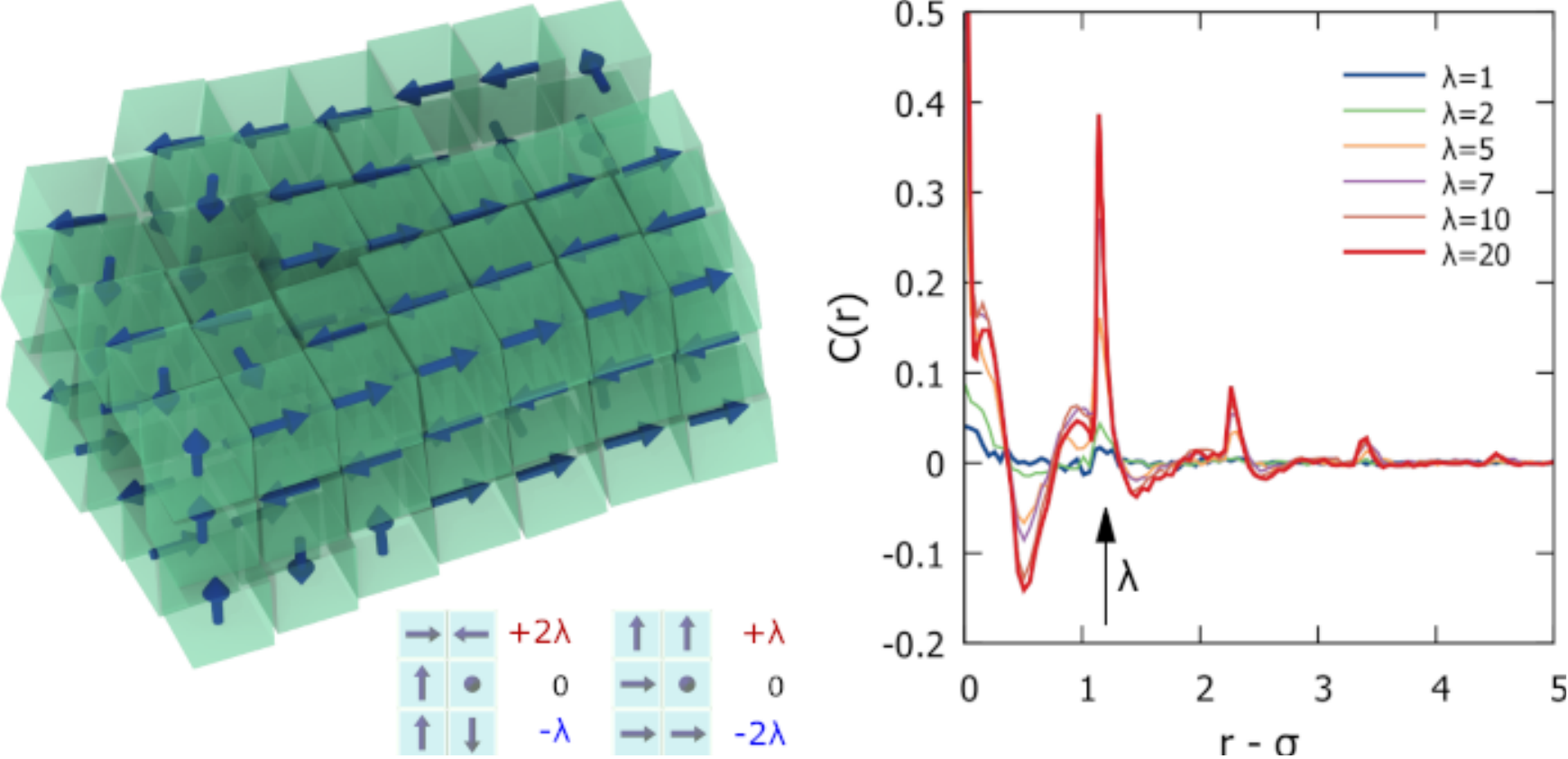}
  \caption{Spin structure of the assembled solids: left - characteristic spin motifs in the assemblies of dipolar nanocubes, right - spin-spin correlation function $C\left(r\right)=\langle \mathbf{s_1}\left(0\right)\cdot\mathbf{s_2}\left(\mathbf{r}\right)\rangle$ at varying dipolar interaction strength showing the presence of mixed ferroelectric-antiferroelectric ordering.}
  \label{fgr:spin}
\end{figure}

\begin{figure}[h]
\centering
    \includegraphics[width=1.0\linewidth]{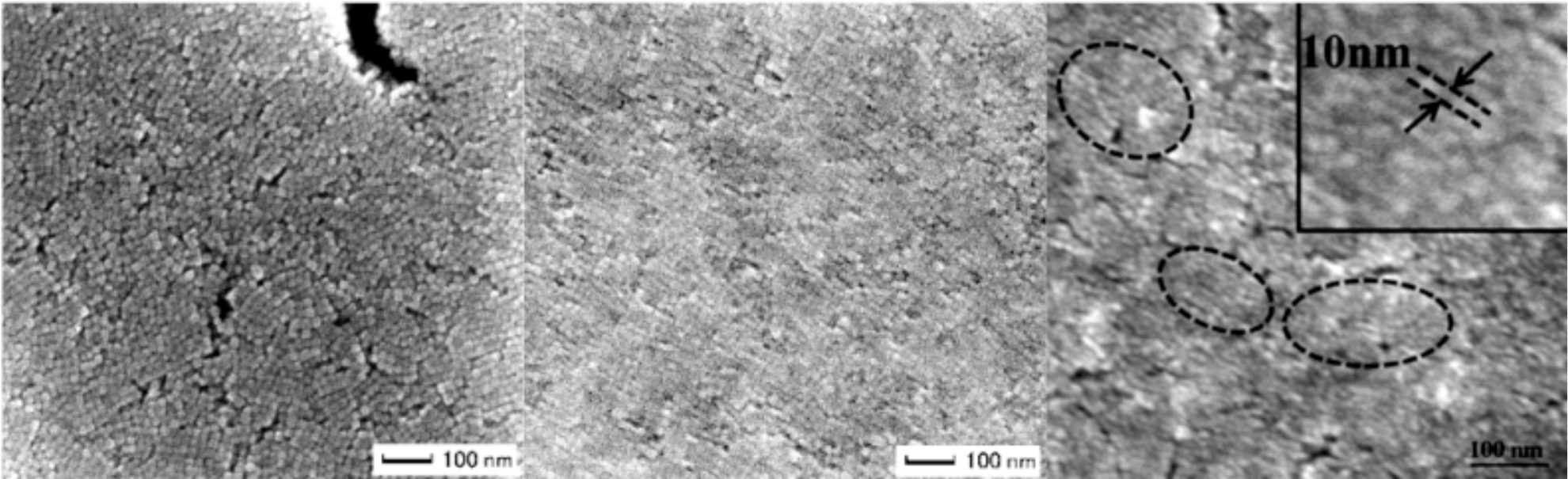}
  \caption{FE-SEM images of BaZr\textsubscript{x}Ti\textsubscript{1-x}O\textsubscript{3} nanocube assemblies: disordered at $x=0.1$ (left) and regular at $x=0.2$ (middle) [reproduced from Ref.~\cite{Mimura2016b}, Copyright (2016), The Japan Society of Applied Physics]; SEM image of the assembled Ba\textsubscript{0.8}Sr\textsubscript{0.2}TiO\textsubscript{3} nanocube film (right) [reproduced from Ref.~\cite{Su2016} with permission from The Royal Society of Chemistry].}
  \label{fgr:exper_sol}
\end{figure}

While the polarization structure of single-domain ferroelectric nanoperovskites is highly ambiguous owing to major influence from many uncontrolled factors, the phase state of their ferromagnetic analogues is rather well understood\cite{wohlfarth1980ferromagnetic}. Hence, we start discussing the experimental implications of dipole-mediated self-assembly using magnetic systems as a reference. In contrast to conventional magnetite Fe\textsubscript{3}O\textsubscript{4}\cite{Singh2014,Wang2018}, which is a relatively soft magnetic material with $\langle 111\rangle$ easy axis, cobalt-substituted ferrite Co\textsubscript{x}Fe\textsubscript{3-x}O\textsubscript{4} ($x > 0.01$)
has high magnetocrystalline anisotropy and $\langle001\rangle$ easy axis\cite{wohlfarth1980ferromagnetic} corresponding to the direction of a spontaneous dipole. Recently, monodisperse single-domain Co\textsubscript{x}Fe\textsubscript{3-x}O\textsubscript{4} (x=0.4-0.5) nanocubes have been accessed by solution phase thermal decomposition reaction\cite{Yang2015} of Co(acac)\textsubscript{2} and Fe(acac)\textsubscript{3} in high boiling point solvent in the presence of oleic acid, which may enable next generation data storage and theranostic applications, but also provide an ideal system for validating our proposed self-assembly scenario. Fig.~\ref{fgr:simul}B shows a scanning electron microscopy (SEM) image of approx. 35 nm nanocubes ($\lambda\gg 1$, Ref.\cite{Zablotsky2017}) self-assembled on the Si substrate by controlled evaporation of the solvent from a hexane solution in good agreement with simulated surface morphology (Fig.~\ref{fgr:simul}A). The spin-spin correlation function (Fig.~\ref{fgr:spin}) describes the dipolar structure of the simulated solid and its fast decay indicates that only the local neighborhood is correlated, whereas the solid itself is in a paraelectric (paramagnetic) state. The local neighborhood is marked by a ferroelectric motif\cite{Takae2018} reflecting the chaining of  individual nanocube dipoles\cite{Zablotsky2017}, which is the most energetically advantageous ($U_{dd}=-2\lambda kT$), with noticeable antiferroelectric inclusions\cite{Talapin2007} due to a side-to-side aggregation of neighboring chains. Fig.~\ref{fgr:spin} shows a characteristic spin structure fragment of a crystallite embedded within a mostly disordered layer.

\subsection{Nanoperovkite cubes}

Experimentally, the room-temperature colloidal assembly of BaTiO\textsubscript{3} nanocubes within an evaporative process varies from disordered assemblies with random internal structure \cite{Mimura2011,Mimura2012c} to partially ordered 3D arrays with a typical domain size of about $\propto$100 nm (approx. 5 particle sizes) \cite{Mimura2011,Mimura2012c}. However, dense multilayered superlattice assemblies of BaTiO\textsubscript{3} nanocubes\cite{Mimura2012,Mimura2012b,Mimura2013,Mimura2015,Mimura2016} were also produced in some experiments, where the kinetic effects have been systematically controlled. The surface and inner structure were highly ordered with SC lattice and a packing fraction up to 90\% \cite{Mimura2012b, Kato2013} over a relatively large area of up to $\propto$ 10 $\mu m^2$ \cite{Mimura2012,Mimura2013,Mimura2015b,Mimura2016} with thickness varying from several hundred nanometers \cite{ Mimura2012b,Mimura2013,Kato2013} up to a $\propto$ 1 $\mu m$\cite{Mimura2015b} (several examples are shown in Fig.~\ref{fgr:exper}A). 
This leads to an important conclusion - despite the evidence discussed above, 20 nm nanocubes of BaTiO\textsubscript{3} must have no apparent dipole or otherwise the assembly of a long-range regular lattice would not have been experimentally feasible. 

It appears that the polar state of nanoperovskites is extremely sensitive to the electrochemical state of the surface \cite{Yang2017,Kalinin2018}. For one, the retention of the nanoscale ferroelectric order crucially relies on the ability of the molecular adsorbates to screen the diverging depolarizing field due to surface charges associated with a spontaneous polarization by inducing an electric field in the opposite direction\cite{Spanier2006,Szwarcman2014,Polking2012}. Specifically, surface hydroxyls (OH) typically adsorbed on metal oxide nanoparticles during the synthetic process and carboxylates (R-COO, e.g. oleate) extensively used to passivate VdW interactions in colloidal processing of perovskites \cite{Mimura2012c,Parizi2014,Parizi2014,Caruntu2015}
 can produce an effective charge compensation as previously shown by quantum chemical calculations\cite{Spanier2006}. 
Hence, dipole-dipole interaction is stronger with weaker presence of adsorbates and incomplete charge screening, whereas the observation of electrostatic fringing fields emanating from 20 nm oleic acid capped BaTiO\textsubscript{3} nanocubes in electron holography experiments \cite{Polking2012} suggests that the magnitude of the dipole moment is an unknown variable with a strong dependence on the delicate nuance of the synthetic route \cite{Polking2012, Szwarcman2014} with a potentially destructive effect in self-assembly experiments.

Recent systematic work done by Mimura and Kato\cite{Mimura2011,Mimura2012,Mimura2012c,Mimura2012b,Mimura2013,Mimura2016b,Mimura2015,Mimura2016,Kato2013} provides strong evidence that the outcome of self-assembly can in fact be significantly affected by the phase state of nanoperovskites. The 20 nm strontium titanate SrTiO\textsubscript{3} nanocubes\cite{Dang2011} could not be assembled and produced a mostly random glassy structure with some embedded SC crystallites (Fig.~\ref{fgr:exper}B, left), whereas a solution of BaTiO\textsubscript{3}-SrTiO\textsubscript{3} nanocubes mixed in equal proportions yielded under an identical procedure a long-ranged regular cubic lattice where both types of nanocubes are homogeneously distributed \cite{Mimura2012, Mimura2012b} (Fig.~\ref{fgr:exper}B, right). Dynamic light scattering (DLS) of SrTiO\textsubscript{3} nanocube solutions has shown that they were slightly aggregated\cite{Mimura2012b}, which is compatible with the chain-fluid state\cite{Zablotsky2017,Zablotsky2019}. Similar equilibrium structures have been imaged by cryo-TEM in vitrified solutions of PbSe and piezoelectric CdSe nanoparticles\cite{Klokkenburg2007}. The analysis based on statistical thermodynamics places the interaction strength at 8$\div$10 $kT$ ($\lambda=8\div10$). 
Previous simulation and experimental studies suggest that dense equilibrated structures of cubes well tolerate particle size\cite{Agarwal2012, Quan2012} and shape\cite{Ni2012, Gantapara2013, Quan2012} imperfections robustly producing cubatic order $P_4$ during self-assembly, whereas a strong enough interaction is sufficient to block the development of long-range orientational order during consolidation (cf. Fig.~\ref{fgr:g4map}).

While bulk SrTiO\textsubscript{3} is an "incipient ferroelectric", which never achieves ferroelectric coherence\cite{Ihlefeld2016}, however, the surface of pristine SrTiO\textsubscript{3} probed by graphene was reported to be in a ferroelectric-like polar state\cite{Sachs2014} having a non-zero static dipole moment, possibly due to an outward displacement of oxygen atoms (surface rumpling \cite{Bickel1989}). 
We did quantum chemical calculations for this system, which indicate that the surface rumpling of "naked" SrTiO\textsubscript{3} could result in an overall $\langle001\rangle$ dipole of approx. 3560 D and $\lambda=22$ (1500 D and $\lambda=4$ within a shell model\cite{Heifets2000}) for a 20 nm nanocube. 
The examination of the pre-edge spectra of x-ray absorption reported that the single phase monodispersed oleic acid-capped SrTiO\textsubscript{3} nanoparticles acquire a significant off-centering of Ti characteristic of a tetragonal distortion in the size range somewhere between 10 and 82 nm entering a room-temperature polar (and possibly ferroelectric) structural state \cite{Tyson2014}, which could explain the disordered/polycrystalline assembly of SrTiO\textsubscript{3} nanocubes. Similarly, it is well known that the dipolar interaction can be screened by coordination with non-dipolar particles in a mixed system. 

Likewise, the phase state of ABO\textsubscript{3} perovskites can be rationally modified by homovalent substitution into the A or B sites. It was found that B -site doping by Zr\textsuperscript{4+} in BaZr\textsubscript{x}Ti\textsubscript{1-x}O\textsubscript{3} perovskite reduces the magnitude and disrupts the coherence of cooperative dipolar order formed by the off-center Ti\textsuperscript{4+} displacements in their Ti-O\textsubscript{6} octahedra\cite{Rabuffetti2013,Acosta2017}. At low Zr content ($x < 0.1$) a normal ferroelectric behavior was shows, whereas further substitution proceeds via a multiphase point at $x\approx0.15$, in which a superposition of rhombohedral, orthorhombic, tetragonal and cubic perovskite phases coexist near room temperature, to a predominantly cubic phase ($x \approx 0.2$). Here the compositional dependence of the room temperature local crystal structure is not affected by size effects down to sub-20 nm BaZr\textsubscript{x}Ti\textsubscript{1-x}O\textsubscript{3} nanocrystals\cite{Rabuffetti2013}.
Indeed, the assemblies of BaZr\textsubscript{x}Ti\textsubscript{1-x}O\textsubscript{3} recently produced by dip-coating\cite{Mimura2016b} (Fig.~\ref{fgr:exper_sol}) did show a strong dependence on Zr\textsubscript{x} content under identical experimental procedure - forming distinct crystalline phases embedded within a largely polycrystalline morphology at $x=0.1$ and a superior long range ordering at a higher Zr substitution ($x=0.2$).
In turn, upon A-site substitution of Ba\textsuperscript{2+} by Sr\textsuperscript{2+} the critical levels necessary to induce a tetragonal-to-cubic phase transition are higher $x \approx 0.30\div0.50$\cite{Rabuffetti2013,Acosta2017} and monodisperse oleic acid/oleylamine-capped 10 nm Ba\textsubscript{0.8}Sr\textsubscript{0.2}TiO\textsubscript{3} nanocubes were assembled by the spin coating procedure into a mostly disordered 300 nm thick film (Fig.~\ref{fgr:exper_sol}, right), where unusual embedded periodical stripe patterns were observed composed of 10 nm nanocubes oriented along the same directions\cite{Su2016}, which is strongly indicative of an anisotropic pairwise interaction.

\section{Conclusions}\label{sec:conclusions}

We have studied for the first time the morphology of densified ensembles of dipolar nanocubes obtained under their osmotic compression. Emerging materials technologies demand superior film deposition strategies that can be absorbed into an entirely solution-based processing compatible with a current array of coating, spraying, (ink-jet) printing or patterning methods using solubilized nanoperovskites. Deterministic assembly via a colloidal route is a core platform for exploiting the potential for variable property design and deployment of mesoscale architectures for energy-harvesting devices and their integration into a successful energy-management concept. Based on their ability to tile space free-standing monodisperse nanocubes are premier candidates for this design paradigm, however, achieving long-range order in self-assembled supercrystals is still challenging. 

Our study indicates that while simple cubes order reliably into regular lattices, nanocrystal cubes interacting via a long-range anisotropic potential go through a different pathway leading to unavoidable polycrystallinity and glassiness, whereas the strength of the dipolar interaction affects the final size of the crystallites embedded in a disordered glassy matrix. This behavior is also very different from the assembly of spherical nanocrystals where ferroelectric and ferromagnetic dipolar coupling should stabilize exotic hcp (hexagonal close-packed) and sh (simple hexagonal) superlattices\cite{Talapin2007}. Our results reproduce the morphologies observed in the experiments and are compatible with a variety of practical osmotic pressure-based techniques starting from a dilute solution. 

Exploring the features of mesoscale disorder induced by competing anisotropies we find characteristically similar behavior in a wide array of systems of different physical nature - both ferromagnetic and ferroelectric. The basic understanding of the spontaneous polarization in soluble ferroelectric monodomains is incomplete and it is difficult to ascertain its presence by direct measurement, whereas the self-assembled phases reflect the structure of pair-wise interactions. We show that in fact for solubilized BaTiO\textsubscript{3} nanocubes the polarization screening must be \textit{complete} without appreciable dipole moment (below a few $\propto kT$), whereas stray interactions from incompletely screened surface charges are likely major sources of disorder in self-assembled architectures of some nanoscale perovskites, most specifically SrTiO\textsubscript{3}.
Similarly, the distortion of the central symmetry in zincblende piezoelectric ZnSe\cite{Shim1999} by the surface electronic states\cite{Noguera2012} produces a large size-dependent permanent dipole. The polar nature of nanocrystals (e.g. rocksalt PbSe \cite{Klokkenburg2007}, zincblende CdTe \cite{Tang2006}, wurtzite piezoelectric CdSe nanocubes \cite{Klokkenburg2007}, transition metal oxides \cite{Pacholski2002,Polleux2004}) driving their anisotropic assembly into a variety of dipolar mesostructures\cite{Tang2002,Klokkenburg2007,Tang2006} ($\lambda\gg 1$\cite{Klokkenburg2007}) may be a universal feature of nanodielectrics \cite{Shim1999,Noguera2012}. Hence, operating the capping ligands to control the surface chemistry\cite{Kuzovkov2011a,Kuzovkov2011b,Kuzovkov2011c,Yang2017} might be central to produce long-range order and will require thorough understanding of the adsorbate induced polarization screening\cite{Mastrikov2018} in future studies.

Understanding the origins of order frustration\cite{Baule2018} in perovskite superlattices is highly desirable for the rational design of ferroelectric devices from elementary building block. Colloidal shape engineering provides access to a class of entropically assembled morphologies encoded solely by the shape anisotropy inducing angularly specific interactions. Harnessing competing or synergistic entropic-enthalpic anisotropies for purposeful valence engineering may lead to a broader class of design strategies in nonclassical crystallization, which could be tuned to target specific morphologies \cite{vanAnders2013, Takae2018}. 

\section{Acknowledgments}
The authors thank Marjeta Ma\v{c}ek Kr\v{z}manc for many useful discussions. The financial support of M-ERA.NET Project HarvEnPiez (Innovative nano-materials and architectures for integrated piezoelectric energy harvesting applications) is gratefully acknowledged. D.Z. acknowledges the support of the postdoctoral research program at the University of Latvia (Project No. 1.1.1.2/VIAA/1/16/072). The computing time of the LASC cluster was provided by the Institute of Solid State Physics (ISSP).


\bibliography{rsc} 
\bibliographystyle{rsc} 

\end{document}